\documentclass[aps, pra, reprint,amsmath,amssymb,superscriptaddress]{revtex4-2}

\usepackage[]{hyperref}
\hypersetup{
    colorlinks=true,
    linkcolor=blue,
    filecolor=magenta,      
    urlcolor=cyan,
    citecolor = blue
    }

\usepackage{graphicx}
\usepackage{dcolumn}
\usepackage{bm}
\usepackage{xcolor}
\usepackage{comment}
\usepackage{bbm}

\newcommand{\diff}{\mathrm{d}}
\newcommand{\avg}[1]{\langle #1 \rangle}

\newcommand{\vect}[1]{\mathbf{#1}}

\newcommand{\affilPhotonics}{Photonics Laboratory, ETH Zürich, 8093 Zürich, Switzerland}
\newcommand{\affilQC}{Quantum Center, ETH Zürich, 8093 Zürich, Switzerland}
\newcommand{\affilPadova}{Università di Padova, Dipartimento di Fisica e Astronomia, 35134 Padova, Italy}
\newcommand{\affilDelft}{Kavli Institute of Nanoscience, Department of Quantum Nanoscience,
TU Delft, 2628CJ Delft, The Netherlands}
\newcommand{\affilIST}{Institute of Science and Technology
Austria, Am Campus 1, 3400 Klosterneuburg, Austria}

\begin{document}

\title{Shot-to-shot noise cancellation for parametric oscillators}

\author{Martynas Skrabulis}
\affiliation{\affilPhotonics}
\affiliation{\affilQC}
\author{Martin Colombano Sosa}
\affiliation{\affilPhotonics}
\affiliation{\affilQC}
\author{Nicola Carlon Zambon}
\altaffiliation[Present address: ]{\affilPadova}
\affiliation{\affilPhotonics}
\affiliation{\affilQC}
\author{Andrei Militaru}
\altaffiliation[Present address: ]{\affilIST}
\affiliation{\affilPhotonics}
\affiliation{\affilQC}
\author{Massimiliano Rossi}
\altaffiliation[Present address: ]{\affilDelft}
\affiliation{\affilPhotonics}
\affiliation{\affilQC}
\author{Lukas Novotny}
\affiliation{\affilPhotonics}
\affiliation{\affilQC}
\author{Martin Frimmer}
\affiliation{\affilPhotonics}
\affiliation{\affilQC}

\begin{abstract}

Powerful approaches to squeeze the motional state of a harmonic oscillator rely on the stepwise modulation of its resonance frequency. 
Such protocols can be limited by forces that vary slowly between experimental runs but are constant during a single experimental shot.
Such shot-to-shot noise gives rise to a spread in experimental outcomes that masks the uncertainty intrinsic to quantum theory.
Taking inspiration from spin-echo protocols, we propose a decoupling technique that, under ideal conditions, perfectly cancels shot-to-shot force noise in squeezing experiments based on parametric modulation. 
We implement the protocol using an optically levitated nanoparticle, where shot-to-shot force noise arises from slowly varying stray fields acting on the charge carried by the particle. Using our oscillator-echo protocol, we demonstrate shot-to-shot noise suppression to the measurement-backaction limit. 

\end{abstract}

\maketitle

\section{Introduction}
Mechanical oscillators in the quantum regime are promising candidates to explore the limits of modern physics~\cite{aspelmeyer_RMP_2014_optomechanicsreview, Croquette2023_advances_AVS,carney_QST_2011_mechanicalsensing}. 
The main challenge hindering operation outside the classical realm is decoherence~\cite{Zurek2007_decoherence}. This process can be thought of as entanglement between the mechanical oscillator and other degrees of freedom which remain unobserved, but whose measurement can, in principle, provide relevant information about the state of the mechanics~\cite{Jacobs_introQM_2006}.
For tethered mechanical oscillators, sophisticated phonon engineering~\cite{MacCabe2020,Engelsen2024} has enabled sufficient suppression of decoherence by the thermal bath to reach the regime of measurement-based quantum control~\cite{Rossi2018}, even at room temperature~\cite{Huang2024}. In this regime, the fluctuations acting on the system are dominated by the quantum noise of the probe light. 

A particularly simple optomechanical platform that is routinely operated in the backaction-dominated regime is an optically levitated nanoparticle~\cite{gonzalez2021levitodynamics, Romero-Isart2011opticallyLevitating, chang2010cavity}. The center-of-mass motion of such a dipolar scatterer trapped in a laser focus resembles a three-dimensional harmonic oscillator whose position is continuously measured. Decoherence due to the thermal bath (constituted by the gas surrounding the nanoparticle) can be suppressed by operating the optical trap in ultra-high vacuum to reach the backaction-dominated regime~\cite{Jain_2016_PRL_Photonrecoil}. 
Quantum control of optically levitated nanoparticles has enabled cooling their center-of-mass motion to the ground state~\cite{delic_science_2020_cavitygs,tebbenjohanns_nature_2021_groundstate,magrini_nature_2021_quantumcontrol,Kamba2022optical,piotrowski2023simultaneous}.

Currently, a prime goal of levitated optomechanics is to expose quantum features of the mechanical state of the particle~\cite{romero2011quantumsuperposition, Wan2016PRLfreeNano, Nimmrichter2025electron-enabled}. First steps have been taken within the realm of Gaussian states by squeezing the fluctuations of one mechanical quadrature of the particle below their zero-point value~\cite{Kamba_Science_2025_Quantumsqueezing,Rossi_PRL_2025_QuantumDeloc}.
These approaches exploit a unique feature of optically levitated systems, whose resonance frequency can be tuned over a wide range by controlling the trapping-laser power. 
This mechanism allows engineering a squeezing operation via parametric modulation of the oscillator eigenfrequency~\cite{Ma1989_squeezing}.

Besides decoherence, another obstacle of little fundamental but significant practical concern is the stability of the experimental conditions~\cite{Llordes_PRL_2024_doublewell,neumeier_PNAS_024_supermario,Pedernales_PRA_2022_matterwaveinter}. 
Many experimental repetitions are required to expose the statistical properties peculiar to quantum mechanics. 
Slow fluctuations of the experimental parameters, termed ``shot-to-shot noise", have challenged the levitated-optomechanics community when creating squeezed states of mechanical motion~\cite{Kamba_Science_2025_Quantumsqueezing, Rossi_PRL_2025_QuantumDeloc}.
Such low-frequency fluctuations are considered as shot-to-shot in nature in the following limit. First, the time between consecutive experimental repetitions (shots) is sufficiently long for the uncontrolled parameter to change appreciably. Second, the duration of a single shot is short enough for the noise to be constant on that timescale.

For example, Kamba \textit{et al}. have identified the difference in the center of optical mass and center of inertial mass together with the particle's orientation as a contributor to shot-to-shot fluctuations~\cite{Kamba2023_revealingVel}. They have averaged this noise to zero during each single realization of the experiment by rapidly spinning the particle~\cite{Kamba_Science_2025_Quantumsqueezing}.
A very different type of shot-to-shot noise is encountered in experiments using a charged particle~\cite{Seta2025_shot2shotInverted}, which experiences a Coulomb force due to uncontrolled electric stray fields.
In practice, the duration of a single experimental shot is much shorter than the temporal spacing between the runs, such that a slowly varying electric stray field manifests as shot-to-shot force noise limiting effective state purity~\cite{Rossi_PRL_2025_QuantumDeloc}. 
Reference~\cite{Rossi_PRL_2025_QuantumDeloc} has taken first steps to significantly reduce the impact of such shot-to-shot force noise by suitably modulating the particle's trapping potential. 

At a more abstract level, shot-to-shot noise observed in optomechanics effectively yields an ensemble broadening, where each member of the ensemble corresponds to one realization of the experiment.
This situation is reminiscent of the effective dephasing encountered by an ensemble of spins whose Larmor frequencies slightly differ~\cite{Slichter_NMRbook}. 
The existence of powerful echo protocols that refocus spin ensembles and effectively cancel the ensemble dephasing~\cite{Hahn1950_spinEchoes, Viola_1998_dynamicalSuppr} raises the question whether similar decoupling techniques exist to make state-expansion protocols for mechanical oscillators robust against shot-to-shot noise.

Here, we propose and experimentally realize a protocol that cancels shot-to-shot force noise in squeezing protocols for harmonic oscillators that rely on a stepwise modulation of the oscillator frequency.
Our oscillator-echo protocol embeds the desired squeezing manipulation in between two decoupling steps that effectively refocus the ensemble akin to a spin-echo protocol. 
We implement our protocol using an optically levitated nanoparticle as an experimental platform and demonstrate shot-to-shot force-noise suppression to the limit where the observed quadrature variances are limited by measurement backaction.

\section{Experimental setup}
\begin{figure}[tb]
\includegraphics[width=0.9\columnwidth]{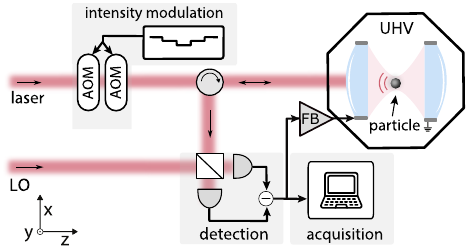}
\caption{\label{fig:setup} 
Experimental setup. A dielectric nanoparticle is optically trapped in a focused laser beam (optical axis $z$) in ultra-high vacuum (UHV). The phase of the light backscattered by the particle is compared to a local oscillator (LO) to measure the particle position along $z$. 
A feedback (FB) force is applied to cool the particle motion. 
The stiffness of the optical trap is modulated via the laser intensity with a pair of acousto-optic modulators (AOMs).}
\end{figure}

Our experimental setup is the one detailed in Refs.~\cite{Rossi_PRL_2025_QuantumDeloc,skrabulis2026nanomechanicalsensorresolvingimpulsive}, and schematically illustrated in Fig.~\ref{fig:setup}. 
A silica nanoparticle (nominal diameter 100 nm, mass 1.2 fg) is trapped in an $x$ polarized laser beam (wavelength 1550~nm, power $\sim$ 600~mW, optical axis $z$) focused with an aspheric lens (numerical aperture~0.8).
The center-of-mass oscillation frequencies of the nanoparticle are $(\Omega_z,\Omega_x,\Omega_y)/(2 \pi) =(52, 141, 175$)~kHz. We focus on the motion along $z$ throughout the remainder.

We use cold-damping to reduce the center-of-mass temperature of the particle to a phonon population of $n = 1.2 \pm 0.6$, limited by the detection efficiency $\eta=0.14$ of our system~\cite{tebbenjohanns_nature_2021_groundstate}. 
The feedback is implemented by acting on the particle (which carries a few elementary charges~\cite{Frimmer2017_netCharge}) with a Coulomb force that is proportional to the particle speed, as derived from the position measurement \cite{tebbenjohanns_PRL_2019_colddamping,Conangla2019_optimalFB}. 
Measurement backaction, i.e., the radiation pressure shot noise exerted by the trapping laser, gives rise to a recoil heating rate $\Gamma_\text{qb}/(2\pi)= \text{3.4}\pm \text{0.5}\,\text{kHz}$.
A pair of acousto-optic modulators allows us to control the trap-laser intensity in a stepwise manner (switching time $\sim200~$ns, much faster than the particle's oscillation period). A reduction of the laser intensity by a factor $r^2$ has a two-fold effect: it decreases the oscillator frequency by $r$ and, simultaneously, reduces the backaction rate by $r^2$.

\section{State evolution during a frequency jump}

One approach to manipulate the quadrature uncertainties of a levitated nanoparticle is to instantaneously switch the oscillator's resonance frequency from the initial value $\Omega$ to the reduced value $\Omega/r$, with $r>1$, for a finite time \cite{Bonvin_PRL_2024_hybridtrap,Rossi_PRL_2025_QuantumDeloc,Kamba_Science_2025_Quantumsqueezing,mattana2025traptotrapfreefallsoptically,steiner2025freeexpansionchargednanoparticle}. This parametric operation is termed a ``frequency-jump'' \cite{Xin_PRL_2021_frequencyjump,Janszky_PRA_1992_strongsqueezing}. 
In the following, we revisit what happens to a Gaussian state under a frequency jump. 

We model a quantum harmonic oscillator of mass $m$ and eigenfrequency $\Omega$ using the state vector $\vect{v}=[Q,P]^T$, where $Q=z/z_\text{zp}$ and $P=p/p_\text{zp}$ are position $z$ and momentum $p$, each normalized by their respective zero-point values $z_\text{zp}=\sqrt{\hbar/(2m\Omega)}$ and $p_\text{zp}=\sqrt{\hbar m\Omega/2}$. 
The evolution of the state vector is governed by the quantum Langevin equation
\begin{equation}\label{eq:stateVectorEvolution}
    \dot{\textbf{v}}(t)=A\textbf{v}(t) + \textbf{w}(t) + \textbf{f}(t).
\end{equation}
The first term describes the deterministic evolution of the state vector in the harmonic potential. For a potential with reduced eigenfrequency $\Omega/r$ the drift matrix reads
\begin{equation}
    A = \begin{pmatrix}0 & \Omega \\ -\Omega/r^2 & 0 \end{pmatrix}    .
\end{equation}
The second term $\vect w(t)$ describes a stochastic force acting on the oscillator, which can, e.g., be due to a thermal bath.
For an optically levitated nanoparticle in ultra-high vacuum, this stochastic force is dominated by measurement backaction according to $\vect{w}(t)=\sqrt{\Gamma_\text{qb}/r^2}X_\text{in}(t)[0,1]^T$, where $X_\text{in}(t)$ is the zero-mean amplitude quadrature of the tweezer field with autocorrelation function $\langle X_\text{in}(t)X_\text{in}(t')\rangle = \delta(t-t')/2$. 
The third term $\vect f(t)=f(t)[0,1]^T$ describes the action of a deterministic force $f(t)$. In the following, we model the effect of shot-to-shot noise by assuming that the force $f(t)=f_0$ is constant over a single run of the experiment but can change from run to run.

Equation~\eqref{eq:stateVectorEvolution} determines the evolution of any initial state vector $\vect v_0 = \vect v(t=0)$ during a time $t$ to the final state 
\begin{equation}\label{eq:generalEvolution_v(t)}
    \textbf{v}(t) = \Phi_r(t)\textbf{v}_0 +\int_{0}^{t}\Phi_r(t-s) [\textbf{w}(s) + \textbf{f}_0] \diff s .
\end{equation}
The state-transition matrix transforming the initial state vector reads
\begin{equation}
    \Phi_r(\theta) = \begin{pmatrix} \cos(\theta) & r \sin(\theta) \\ -\sin(\theta)/r & \cos(\theta)\end{pmatrix},
\end{equation}
where $\theta=\Omega t/r$ represents the phase picked up by the oscillator during time $t$. 
Note that $\Phi_r(t)$ describes rotations of phase-space points around the origin on elliptical trajectories. This situation becomes obvious when rewriting the state-transition matrix as
\begin{equation}
    \Phi_r(\theta) = S(\sqrt{r})\Phi_1(\theta)S^{-1}(\sqrt{r}),
\end{equation}
where $\Phi_1(\theta)$ is a simple (circular) rotation matrix, and $S(\sqrt{r})=\mathrm{diag}[\sqrt{r},1/\sqrt{r}]$ describes a squeezing transformation along the phase-space coordinates with squeezing factor $\sqrt{r}$. 

For Gaussian states, the phase-space distribution is fully described by the state's expectation value $\vect d=\avg{\vect v}$ and its covariance matrix
$\Sigma=\langle\textbf{v}\textbf{v}^T\rangle-\langle\textbf{v}\rangle\langle\textbf{v}\rangle^T$.

\subsection{Evolution of expectation values} Let us discuss the evolution of the mean $\vect d$, starting at $\vect d_0=\avg{\vect v_0}$, as given by Eq.~\eqref{eq:stateVectorEvolution}. 
Note that, from here on, we measure Q relative to the minimum of the effective potential of stiffness $\Omega$ in the presence of $f_0$.
Using $\avg{\vect w(t)}=0$, we find
\begin{equation}\label{eq:evolutionMean}
    \vect d(\theta) = \Phi_r(\theta)\vect d_0  + \vect d_{f_0,r}(\theta).
\end{equation}
Therefore, the initial state undergoes a (squeezed) rotation around the origin superimposed on a linear translation by the displacement vector
\begin{equation}\label{eq:evolutionMean_translationVector}
    \vect d_{f_0,r}(\theta)=\frac{f_0}{\Omega} \frac{r^2-1}{r}[r(1-\cos\theta),\sin \theta]^T,
\end{equation}
which is given by the constant force $f_0$, the time of evolution $t$, and the oscillator parameters $\Omega$ and $r$. 

\begin{figure}[t]
\centering
\includegraphics[width=0.9\columnwidth]{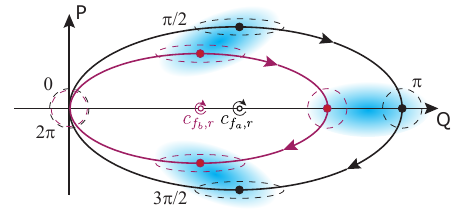}
\caption{\label{fig:principle} 
State-space evolution of a harmonic oscillator in a potential with reduced eigenfrequency $\Omega/r$ for two different values of the force $f_a$ (black) and $f_b<f_a$ (red). 
The oscillator is initialized to a circular Gaussian state at the origin (dashed circles).
The trajectories of the mean around the rotation centers $\vect{c}_{f_a,r}$ and $\vect{c}_{f_b,r}$ are shown as the solid ellipses. 
The covariances are illustrated as dashed ellipses. After evolution by a phase $\theta=\pi/2$ and $3\pi/2$, the state is maximally squeezed along the momentum axis. 
Averaging over many realizations including $f_a$ and $f_b$ leads to an additional contribution to the effective covariance, illustrated by the blue shading. 
}  
\end{figure}

\begin{figure*}[bt]
\centering
\includegraphics[width=0.8\textwidth]{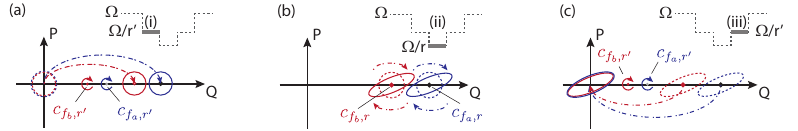}
\caption{\label{fig:principle_echoProtocol} 
State-space evolution during the oscillator-echo protocol. 
(a)~Step (i): The oscillator frequency is instantaneously reduced from $\Omega$ to $\Omega/r'$ for a time corresponding to $\theta_1=\pi$ (see inset for frequency-jump sequence). The state's mean is displaced along the $Q$ axis due to rotation around the center point $\vect{c}_{f_0,r'}$, illustrated for two values $f_a$ (blue) and $f_b$ (red) of the force $f_0$. The state's covariance at the end of the step (solid circle) equals the one at the beginning (dashed) besides an increase due to white-noise heating (not illustrated). 
(b)~Step (ii): The oscillator frequency is reduced further to $\Omega/r$ for any desired time (corresponding to the phase $\theta_2$), e.g., to prepare a squeezed state. 
When $r'$ is chosen correctly relative to $r$, the rotation points during the second step $\vect{c}_{f_a,r}$ and $\vect{c}_{f_b,r}$ fall at the position where the state's mean is located, such that no displacement of the mean occurs during this step.
(c)~Step (iii): The oscillator frequency is set back to $\Omega/r'$ for a time corresponding to $\theta_3=\pi$ as in step~(i). The state is displaced back to the origin, irrespective of the value of the force $f_0$. 
}  
\end{figure*}

It is particularly instructive to visualize the trajectory of $\vect{d}(\theta)$ when starting at the origin $\vect{d}_0=[0,0]^T$.
In this case, the state's mean is given by $\vect{d}_{f_0,r}(\theta)$ and traces an elliptical trajectory around the center point
\begin{equation}\label{eq:rotationPoint}
    \vect c_{f_0,r} = \left[\frac{f_0}{\Omega}(r^2-1), 0 \right]^T.
\end{equation}
This oscillation is intuitively explained as follows. If the potential of stiffness $\Omega$ in the presence of the force $f_0$ has its minimum located at $Q=0$, it will shift to $Q_{f_0,r}=\frac{f_0}{\Omega}(r^2-1)$ when the stiffness is reduced by $r$. Rotation in phase space around the point $\vect c_{f_0,r}$ thus corresponds to oscillations around this new minimum.
In Fig.~\ref{fig:principle}, we illustrate the elliptical trajectories of the mean for two values of the constant force $f_0$.

\subsection{Evolution of covariance matrix}\label{sec:evolutionCovariance}
We now turn to the covariance of the state evolving under a frequency jump according to Eq.~\eqref{eq:generalEvolution_v(t)}. For an initial covariance $\Sigma_0=\Sigma(t=0)$, we find
\begin{equation}\label{eq:covarianceMatrix_sum}
    \Sigma(\theta) = \Phi_r(\theta)\Sigma_0\Phi(\theta)^T + \Sigma_{r}^\Gamma(\theta).
\end{equation}
The first term describes periodic squeezing (and anti-squeezing) of the quadrature variances due to the coherent evolution of the state in the potential with reduced frequency $\Omega/r$. The second term,
\begin{equation}\label{whiteNoiseContribution}
    \Sigma_{r}^\Gamma(\theta)= \begin{pmatrix} \frac{2r\Gamma}{\Omega}\big[\theta-\frac{1}{2}\sin(2\theta)\big] & \frac{\Gamma}{\Omega}\big[1-\cos(2\theta)\big] \\ \frac{\Gamma}{\Omega}\big[1-\cos(2\theta)\big] & \frac{2\Gamma}{r\Omega}\big[\theta+\frac{1}{2}\sin(2\theta)\big] \end{pmatrix},
\end{equation}
describes the monotonous growth of the position and momentum variances due to the white force noise $\vect w(t)$ at the heating rate $\Gamma$. When heating is dominated by backaction, $\Gamma$ equals $\Gamma_\text{qb}$.
We illustrate the coherent evolution of the covariance [first term in Eq.~\eqref{eq:covarianceMatrix_sum}] as the dashed ellipses in Fig.~\ref{fig:principle}. The initially circular state is maximally momentum-squeezed when the phase acquired during evolution in the reduced-stiffness potential is $\pi/2$, and the covariance of the original state is restored at phase $\pi$. 

Figure~\ref{fig:principle} also illustrates the problem arising from a force $f_0$ that is constant during a single iteration of the experiment, but randomly varies from iteration to iteration (shot-to-shot noise). This situation corresponds to averaging over different ellipsoidal orbits of the expectation values, each with a different rotation center (two such orbits are illustrated in Fig.~\ref{fig:principle}). The additional covariance generated by shot-to-shot force noise (illustrated by the blue shading in Fig.~\ref{fig:principle}) is the covariance of the displacement vector in Eq.~\eqref{eq:evolutionMean_translationVector}, and reads
\begin{equation}\label{eq:Sigma_contribution_f_0}
    \Sigma_{f_0,r}^\text{ss} =  \frac{\sigma_{f_0}^2(r^2-1)^2}{r\Omega^2}
    \begin{pmatrix} 
r(1-\cos \theta)^2 & (1-\cos \theta) \sin \theta \\
(1-\cos \theta) \sin \theta & \sin^2 \theta/r 
\end{pmatrix} .
\end{equation}
Here, $\sigma_{f_0}^2$ is the variance of the force $f_0$ evaluated over different iterations of the protocol.
The added variance due to shot-to-shot noise reaches a maximum at half a period ($\theta=\pi$) and vanishes after a full period ($\theta=2\pi$) of evolution (see Fig.~\ref{fig:principle}). 
Therefore, it represents an obstacle for the generation of squeezed states using a frequency jump, which requires an evolution by $\theta=\pi/2$.

\section{Oscillator-echo protocol}

\begin{figure*}[t]
\includegraphics[width=\textwidth]{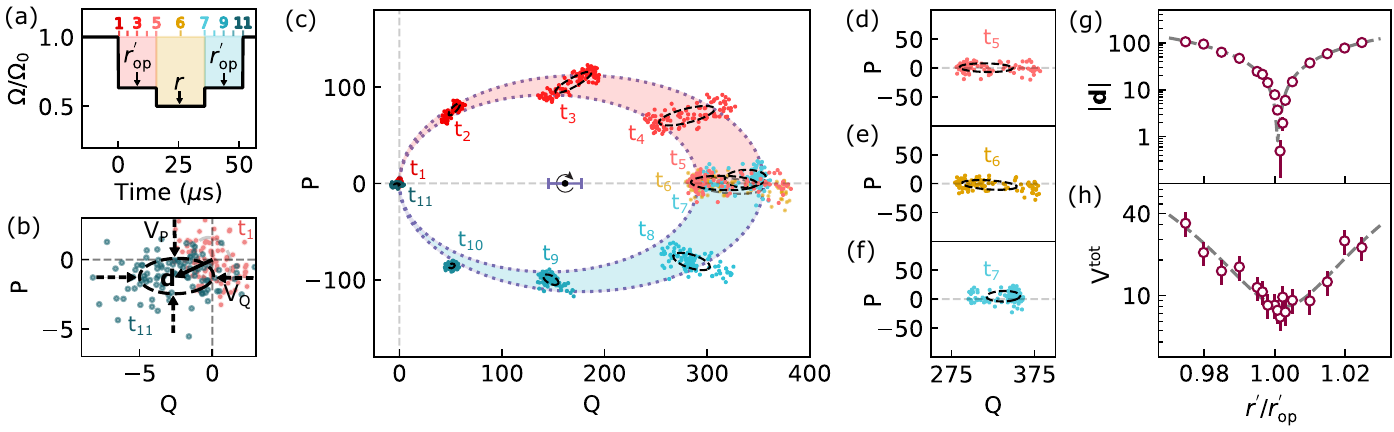}
\caption{\label{fig:echoPulseProtocol} 
(a)~Oscillator frequency during the oscillator-echo protocol realized in panels (b--f) with $r=\sqrt{4}$ and $\Omega/2\pi=52$ kHz. The colored numbers indicate the sampling times at which the state was probed in (b--f).
(b)~Zoom of panel~(c) around the origin of phase-space, showing the state at the beginning (red) and at the end (blue) of the protocol. 
The displacement $\textbf{d}$ of the state's mean by the protocol is shown as the solid arrow. The eigenvalues of the covariance matrix $V_P$ and $V_Q$ represent the main axes of the covariance ellipse (black dashed). 
(c) Evolution of the state throughout the oscillator-echo protocol. 
The color-coding of the datapoints corresponds to the sampling times indicated in panel (a).
The small dashed ellipses (black) illustrate the covariances of the state at the different sampling times. 
The two purple dashed ellipses indicate the trajectories in phase-space as given by stray field values of $f_0\pm\sigma_{f_0}$. 
(d--f)~Zoom of plot (c), showing the state at $t_5, t_6$, and $t_7$, respectively. 
(g)~Measured magnitude of the phase-space displacement $|\vect{d}|$ as a function of $r'$ at the end of the protocol ($t_{11})$. 
(h)~Measured determinant of the covariance matrix $V^\text{tot}=\det(\Sigma)$ at the end of the protocol ($t_{11})$. 
In (g) and (h), the squeezing ratio is $r=\sqrt{10}$ and the dashed gray lines are fits to Eqs.~\eqref{eq:mean_echoPulse} and~\eqref{eq:cov_echoPulse}. See main text for details.}  
\end{figure*}

Can we realize a squeezing protocol using a frequency jump and, at the same time, cancel the contribution to the covariance from shot-to-shot force fluctuations in a way other than ensuring $\sigma_{f_0}^2=0$?
We answer this question affirmatively with a decoupling protocol, which we term the \emph{oscillator-echo protocol}. We illustrate the state evolution throughout our protocol in Fig.~\ref{fig:principle_echoProtocol}. We first provide an intuitive explanation before treating the protocol with full quantitative rigor.

Our protocol consists of three steps. We start with a state initialized at the phase-space origin.
In step (i), the oscillator frequency is instantaneously lowered from $\Omega$ to $\Omega/r'$ for a duration corresponding to the phase angle $\theta_1=\pi$. 
In the presence of a force $f_0$, a state starting at the origin is displaced in phase space along the $Q$ axis due to a rotation around the point $\vect{c}_{f_0,r'}$ given by Eq.~\eqref{eq:rotationPoint}.
The covariance of the state after evolution by $\theta_1=\pi$ returns to its starting value besides the contribution by white-noise heating [see Eq.~\eqref{eq:covarianceMatrix_sum}].
We illustrate the evolution for two different values of the force $f_0$ in Fig.~\ref{fig:principle_echoProtocol}(a). 
In step (ii), illustrated in Fig.~\ref{fig:principle_echoProtocol}(b), the frequency is lowered further to $\Omega/r$ (with $r>r'$) for a time corresponding to the angle $\theta_2$, which can be freely chosen.
Importantly, as we rigorously show below, when picking $r'$ correctly relative to $r$, the rotation center $\vect{c}_{f_0,r}$, during step (ii) falls exactly at the position where the state's mean has been brought to during step (i), irrespective of the value of $f_0$. In other words, step (ii) effects a squeezing operation by a frequency jump without any displacement of the state. 
In step (iii), illustrated in Fig.~\ref{fig:principle_echoProtocol}(c), the frequency is returned to the intermediate value $\Omega/r'$ to acquire the phase $\theta_3=\pi$, before it is switched to the starting value $\Omega$. 
During this last step, the state rotates around the same point as during step (i), which depends on the value of $f_0$, such that the state's mean returns to the starting point, irrespective of the value of $f_0$.

Let us quantitatively model the oscillator-echo protocol. The evolution of the state's mean during the three steps is determined by concatenating the transformations in Eqs.~\eqref{eq:evolutionMean} and \eqref{eq:evolutionMean_translationVector} with the appropriate phase angles and squeezing factors. Noting that $\Phi_{r'}(\pi)=-\mathbbm{1}$, with $\mathbbm{1}$ the identity matrix, we find for the mean of the state at the end of the three-step sequence
\begin{equation}\label{eq:mean_echoPulse}
\begin{split}
    \vect{d}_{r'rr'}(\theta_2) = \Phi_r(\theta_2)\vect d_0 - \Phi_r(\theta_2)\vect d_{f_0,r'}(\pi) \\
    - \vect d_{f_0,r}(\theta_2) + \vect d_{f_0,r'}(\pi).
\end{split}
\end{equation}
Tracing the evolution of the covariance through the three steps using Eqs.~\eqref{eq:covarianceMatrix_sum} and \eqref{whiteNoiseContribution}, we find
\begin{equation}\label{eq:cov_echoPulse}
\begin{split}
    \Sigma_{r'rr'}(\theta_2) =  
    \Phi_r(\theta_2) \Sigma_0 \Phi_r^T(\theta_2) 
    + \Phi_r(\theta_2) \Sigma_{r'}^\Gamma(\pi) \Phi_r^T(\theta_2) 
    \\ +\Sigma_r^\Gamma(\theta_2)
    + \Sigma_{r'}^\Gamma(\pi)
    + \langle\vect{d}_{r'rr'}\vect{d}_{r'rr'}^T\rangle.
\end{split}
\end{equation}
The first term on the right hand side describes the evolution of the initial covariance $\Sigma_0$ during step (ii). Note that steps (i) and (iii)  represent an identity transformation that leaves $\Sigma_0$ unchanged. 
The second term describes the contribution by heating during step (i) according to Eq.~\eqref{whiteNoiseContribution}, propagated through step (ii) without a further effect by step (iii), since it is the identity operation. 
The third term stems from heating during step (ii), unchanged by step (iii), and the fourth term is heating during step (iii). 
The final term resembles the added covariance due to shot-to-shot noise.

Importantly, upon choosing the frequency ratio $r'$ for steps (i) and (iii) to be
\begin{equation}\label{eq:rprime_opt}
    r'_\text{op}=\sqrt{(r^2+1)/2},
\end{equation}
the last three terms in Eq.~\eqref{eq:mean_echoPulse} vanish and the mean evolves according to
\begin{equation}\label{eq:mean_echoPulse_simplified}
    \vect{d}_{r'_{\text{op}}rr'_\text{op}}(\theta_2)=\Phi_r(\theta_2)\vect{d}_0.
\end{equation}
Thus, under this optimal choice $r'_\text{op}$, which only depends on the experimental parameter $r$, the evolution of the mean is entirely determined by step (ii) via $\Phi_r(\theta_2)$. Most importantly, since the evolution of the mean does not depend on the force $f_0$, the last term in Eq.~\eqref{eq:cov_echoPulse} vanishes. As a result, the oscillator-echo protocol cancels any contribution to the final state variance by shot-to-shot fluctuations of the force $f_0$. Notably, the squeezing ratio $r$ and angle $\theta_2$ during step (ii) can be freely chosen.

The price paid for decoupling the squeezing operation in step (ii) from shot-to-shot noise is the heating picked up during the decoupling steps (i) and (iii), represented by the second and fourth term in Eq.~\eqref{eq:cov_echoPulse}.
This additional covariance, in general, depends on $\theta_2$ and $r$. It is instructive to consider an identity operation ($\theta_2=\pi$) during step~(ii). In this case, Eq.~\eqref{eq:cov_echoPulse} simplifies to
\begin{equation}
    \label{eq:covarianceTotalSqueezing}
    \Sigma_{r'_\text{op} r r'_\text{op}}(\pi) = \Sigma_0  + \Sigma_r^\Gamma(\pi) + 2\Sigma_{r'}^\Gamma(\pi),
\end{equation}
where the last term on the right hand side is the penalty due to the decoupling steps.

\section{Experimental demonstration}

We experimentally demonstrate the oscillator-echo protocol in Fig.~\ref{fig:echoPulseProtocol}. 
For the sake of example, we implement an (ideally) identity operation in step~(ii), corresponding to a phase angle $\theta_2=\pi$. For the data shown in Fig.~\ref{fig:echoPulseProtocol}(a--f), we lower the resonance frequency of the oscillator by $r=\sqrt{4}$, such that the phase angle $\pi$ corresponds to the time $t=19$~$\mu$s in the potential softened to $\Omega/r$. 
We embed step (ii) between two frequency jumps by $r'_\text{op}$ according to Eq.~\eqref{eq:rprime_opt}, corresponding to steps (i) and (iii) of the protocol. The full frequency-jump sequence is shown in Fig.~\ref{fig:echoPulseProtocol}(a). 

To characterize the evolution of the state throughout the protocol, we interrupt it before completion by switching the oscillator frequency back to $\Omega$ to sample the phase-space distribution using the method established in Ref.~\cite{Rossi_PRL_2025_QuantumDeloc}. 
The eleven sampling times are indicated in Fig.~\ref{fig:echoPulseProtocol}(a) and labeled $t_s$ with $s\in\left\{1,\ldots, 11\right\}$.
For each sampling time, we repeat the protocol 100 times and display the obtained phase-space coordinates as datapoints in Fig.~\ref{fig:echoPulseProtocol}(c). The point clouds are color-coded to provide the assignment to the sampling times in Fig.~\ref{fig:echoPulseProtocol}(a). 

When observing the point clouds representing the state during step (i) of the protocol (sampling times $t_1$ to $t_5$, corresponding to $0$, $\pi/4$, $\pi/2$, $3\pi/4$ and $\pi$, respectively), we see the mean of the state rotating away from the phase-space origin on an elliptical orbit. 
The center of this orbit [marked with a black circle in Fig.~\ref{fig:echoPulseProtocol}(c)] represents the mean of the force $f_0$ acting on the levitated particle according to Eq.~\eqref{eq:rotationPoint}.
The ellipticity of the orbit is given by $r'$, the squeezing factor applied during steps (i) and (iii) of the protocol.

Regarding the spread of the datapoints, quantified by their covariance matrix [illustrated as the dashed black ellipses in Fig.~\ref{fig:echoPulseProtocol}(c)], we observe an increase during this first step of the protocol, in particular of the position variance. 
Recall that, for a force $f_0$ that is constant over all experimental runs, the covariance matrix after an evolution by $\pi$, i.e., at the end of step (i), returns to its starting value (besides a small contribution due to heating), as described by Eq.~\eqref{eq:covarianceMatrix_sum}. 
Therefore, the observed increase of the covariance ellipse at time $t_5$ [shown separately in Fig.~\ref{fig:echoPulseProtocol}(d)] as compared to $t_1$ is a result of the force $f_0$ changing from shot to shot.

In Figs.~\ref{fig:echoPulseProtocol}(d--f), we show the state at the beginning ($t_5$), in the middle ($t_6$), and at the end ($t_7$) of step (ii). 
The identity operation (phase $\theta_2=\pi$) effected during this step can be thought of as a squeezing step (during the first $\pi/2$, finished at $t_6$) directly followed by an anti-squeezing step (during the second $\pi/2$, completed at $t_7$). 
One would therefore expect the position variance to be boosted by $r^2=4$ at $t_6$ as compared to its value at $t_5$ and $t_7$, while the momentum variance should be suppressed by the corresponding inverse factor $r^{-2}$. 
Strikingly, the effect of the shot-to-shot force fluctuations is strong enough to completely obscure the visibility of any squeezing when observing the state at $t_6$ and comparing to $t_5$ and $t_7$.

The evolution of the state during step (iii) is visualized by the distributions at times $t_8$ to $t_{11}$ in Fig.~\ref{fig:echoPulseProtocol}(c) [corresponding to the phases $\pi/4$, $\pi/2$, $3\pi/4$ and $\pi$, respectively, picked up during step (iii)]. 
Importantly, at the end of the step, the state's mean and variance return close to their starting values, such that the oscillator-echo protocol indeed cancels the effect of the varying force, as intended. 
From our data, we now estimate the fluctuations of the force $f_0$ from shot to shot. 
From the measured covariance matrices at the times $t_1$ to $t_5$, we extract the standard deviation $\sigma_{f_0}=(22.7\pm3.2)~\text{aN}/p_\text{zp}$ of the force $f_0$ using Eq.~\eqref{eq:Sigma_contribution_f_0}. Since the fluctuations of $f_0$ give rise to a spread in rotation centers in phase space, we illustrate $\sigma_{f_0}$ in Fig.~\ref{fig:echoPulseProtocol}(c)  as horizontal errorbars (drawn in purple) on the mean rotation point (black dot).  
Furthermore, we plot the two phase-space trajectories corresponding to $f_0\pm\sigma_{f_0}$ as purple dashed ellipses in Fig.~\ref{fig:echoPulseProtocol}(c). 
As an aside, we remark that it will be interesting to investigate higher-order moments of $f_0$ in future work. 

We now turn to a quantitative assessment of the noise cancellation provided by the oscillator-echo protocol.
To this end, we compare the state variance at the beginning of the protocol ($t_1$) and at the end ($t_{11}$). The corresponding data are shown in magnified form in Fig.~\ref{fig:echoPulseProtocol}(b). 
As a measure for the total state size, we evaluate the determinant of the covariance matrix $V^\text{tot}=\text{det}[\Sigma]$. As an illustrative example, for a state whose covariance ellipse is aligned with the $Q$ and $P$ axes, this total covariance is simply the product of the position and momentum variance $V^\text{tot}=\sqrt{V_P V_Q}$, as illustrated in Fig.~\ref{fig:echoPulseProtocol}(b).
The extracted variance of the state at $t_1$ is $V^\text{tot}_{t_1}=4$, and grows to $V^\text{tot}_{t_{11}}=6.5$ at $t_{11}$. The final variance predicted by our model for the ideal protocol is $V^\text{tot,th}_{t_{11}}=5.5$. The discrepancy between experiment and theory can be explained by two contributions. First, our protocol only cancels shot-to-shot noise. In other words, any fluctuation of the force $f_0$ during a single iteration of the experiment remains uncompensated. Second, an additional variance can arise from an imperfection in the execution of the oscillator-echo protocol, such as an imperfectly chosen value of $r'$.  

A hint toward such an imperfect choice of $r'$ is given in Fig.~\ref{fig:echoPulseProtocol}(b), where we observe that the expectation value of the final state does not perfectly return to that of the initial state. We denote this displacement in phase space by the vector $\vect{d}$.  
In Fig.~\ref{fig:echoPulseProtocol}(g), we investigate the magnitude of the displacement $|\vect{d}|$. In this experiment, we use a squeezing ratio $r=\sqrt{10}$.
We sweep the squeezing factor $r'$ of steps (i) and (iii) while appropriately adjusting the duration of both steps to maintain the acquired phase $\theta_1=\theta_3=\pi$. 
We observe that $|\textbf{d}|$ is minimized when $r'$ approaches $r'_\text{op}$, as expected. We fit the data to Eq.~\eqref{eq:mean_echoPulse}, shown as the dashed line, and obtain the fit parameter $f_0= (144\pm17)$~aN$/p_\text{zp}$. 
The minimal value of $|\textbf{d}|$ is ultimately limited in practice by the frequency stability of the oscillator.

Finally, we investigate how the covariance of the state $V^\text{tot}$ is affected by $r'$ in Fig.~\ref{fig:echoPulseProtocol}(h). 
The measured datapoints show a minimum for $r'=r'_{\text{op}}$, as expected. 
We show the fit of Eq.~\eqref{eq:cov_echoPulse} to the data as the dashed line, and extract as a fit parameter $\sigma_{f_0}= (6.95 \pm 0.91)$~aN$/p_\text{zp}$. 
The difference of the values of $\sigma_{f_0}$ extracted from Figs.~\ref{fig:echoPulseProtocol}(c) and (h) can be explained by the use of a different particle with a different charge-to-mass ratio.
Remarkably, the values found for $\sigma_{f_0}$ closely align with the force fluctuations observed in a different experimental setup reported in Ref.~\cite{Seta2025_shot2shotInverted}. 
It remains to be explored whether this match hints at the mechanism by which the electric stray-field fluctuations are generated.
 
The minimal variance achieved experimentally in Fig.~\ref{fig:echoPulseProtocol}(h) according to the fit is $V^\text{tot}=8.03\pm 1.07$. This result matches the expectation $V^\text{tot}_\text{th}=7.9$ calculated with Eq.~\eqref{eq:covarianceTotalSqueezing} using the experimentally determined values of $\Gamma_\text{qb}$, $n$, and $\eta$ quoted in the experimental section of this work. 
Our measurements therefore demonstrate suppression of shot-to-shot noise to a level where white-noise heating (dominated by measurement backaction in our experiment) is the main contributor to the total state variance, as spelled out in Eq.~\eqref{eq:covarianceTotalSqueezing}.
In summary, Figs.~\ref{fig:echoPulseProtocol}(g) and (h) show that, indeed, there exists an optimal value of $r'$ which minimizes the residual displacement of the state's mean, and reduces the state's variance to the backaction limit. These findings further underline our solid understanding of the experimental system, and the effectiveness of the oscillator-echo protocol for noise cancellation.

\section{Conclusions}

In conclusion, we have proposed and demonstrated an experimental technique to suppress the detrimental effects of shot-to-shot force fluctuations in levitated-optomechanics squeezing experiments. 
Our protocol relies on embedding the state-expansion step of the protocol between two decoupling steps which together cancel the effects of the force before and after the expansion step.
We have experimentally demonstrated the cancellation of shot-to-shot force noise to the limit posed by the backaction noise present during the additional decoupling steps required by the protocol.

We envision that our oscillator-echo protocol will be useful in future levitodynamics experiments that aim to generate extremely squeezed states of mechanical motion \cite{marocco2026threedimensionalsqueezingopticallylevitated}. 
Numerous proposals rely on a charged particle to generate non-Gaussian states of motion, for example by interfacing the nanoparticle with anharmonic electric potentials or to enable long free evolutions in the presence of gravity~\cite{Llordes_PRL_2024_doublewell,Casulleras_PRA_2024_nonGaussian,neumeier_PNAS_024_supermario}. 
Our protocol will aid in protecting those experiments from the detrimental effects of inevitable electric stray fields~\cite{brownnutt2015ion,teller2021heating}. 
Furthermore, it will be interesting to extend the protocol to related scenarios, such as evolution in an inverted potential \cite{Tomassi_PRR_2025_acceleratedexpansion, Seta2025_shot2shotInverted}.

In addition, our protocol will enable further improvements in impulsive-force sensing with levitated nanoparticles~\cite{skrabulis2026nanomechanicalsensorresolvingimpulsive,marocco2026threedimensionalsqueezingopticallylevitated}, with promising applications in identifying dark matter and testing theories that predict elementary particles outside the current standard model~\cite{moore_QST_2021_newphysics, Carney_PRXQ_2023_neutrinosearch,Monteiro_PRL_2020_compDM, carney_QST_2011_mechanicalsensing}.

Finally, our protocol is a harmonic-oscillator analogy of spin-refocusing and dynamical decoupling techniques~\cite{Slichter_NMRbook}, such as the Hahn echo~\cite{Hahn1950_spinEchoes}, developed in the context of nuclear magnetic resonance and electron spin resonance~\cite{Viola_1998_dynamicalSuppr}. 
Therefore, this work provides a bridge between these foundational spin-control techniques and modern nano-mechanical oscillators. This connection may inspire the development of further techniques to combat noise and experimental instabilities in optomechanics.

\subsection*{Acknowledgments}

This research has been supported by the Swiss SERI Quantum Initiative (grant no. UeM019-2), the Swiss National Science Foundation (grant no. 51NF40-160591), and the European Research Council (ERC) under the grant agreement No. [951234] (Q-Xtreme ERC-2020-SyG). M.C.S. thanks for support through an SNSF Fellowship (grant no. 224465). 

\subsection*{Data availability statement}
The data illustrated in the figures of this manuscript will be made publicly available at the time of publication.

\bibliography{bibliography_skrabulis}

\end{document}